\documentstyle[12pt,times,aaspp4]{article}

\lefthead{Black, D.C.}
\righthead{Brown Dwarfs}

\begin{document}

\title{Possible Observational Criteria for Distinguishing\\
    Brown Dwarfs from Planets}

\author{David C. Black}
\affil{Lunar and Planetary Institute, Houston, TX   77058}


\def\lacute{\mathopen{<}}
\def\racute{\mathopen{>}}
\def\lapprox{$_<\atop{^\sim}$}
\def\gapprox{$_>\atop{^\sim}$}


\begin{abstract}
The difference in formation process between binary stars and 
planetary systems is reflected in their composition as well as 
their orbital architecture, particularly orbital eccentricity as 
a function of orbital period.  It is suggested here that this 
difference can be used as an observational criterion to distinguish 
between brown dwarfs and planets.  Application of the orbital 
criterion suggests that with three possible exceptions, all of the
recently-discovered substellar companions discovered to date may 
be brown dwarfs and not planets. These criterion may be used as 
a guide for interpretation of the nature of sub-stellar mass 
companions to stars in the future.

\end{abstract}




%

\section{Introduction}

It is not unusual in science to find that the initial early detection 
of a phenomenon is followed by a rapidly increasing discovery rate as 
interest intensifies and new technology is developed.  Such has been 
the case with the search for substellar companions to stars in the 
solar neighborhood.  While there have been claims of detections of 
companions to stars with masses less than the lower limit for the
mass of a star\footnote[1]{Roughly 80 Jupiter masses.  Normally 
these masses are expressed in terms of solar mases, but the mass 
of Jupiter is a more appropriate unit for this discussion}, most have not stood the test of time.  However, 
that has changed following the announcement of the detection of a 
companion to the star HD 114762 (Latham et al.  1989).  During the 
interval of time from 1995 to early 1997 there have been reported 
detections of 20 substellar mass companions to nearby stars.
 
The recent avalanche of detections began with the paper announcing the 
discovery of a companion to the star 51 Pegasi (Mayor and Queloz 1995).  
The lower limit to the mass of the companion is 0.45 Jupiter masses.  
More remarkable than the mass of the companion was its orbital period, 
4.23 days.  The semi-major axis of this companion is only 0.05 AU!  
The detection of the radial velocity variation for 51 Peg was confirmed 
and there began a rapid sequence of detections of what have been called 
by their discoverers ``extrasolar planets''.

But, are these companions really planets?  On what basis was that 
interpretation of the data made, and how firm is it?  Could these objects 
be something else, like the bottom end of the star formation process, 
viz., brown dwarfs, or an as yet unidentified class of astronomical 
object?  Indeed, there now is a serious challenge (Gray 1997) to the
interpretation of the 51 Pegasi signal as being due to a companion of 
any kind.  It may be that all of the short period signals that have been 
attributed to planetary companions may be intrinsic to the star, but 
that will become clear with additional observations of the type conducted 
by Gray.  The criteria suggested here do not depend upon the reality of those
companion systems.

The fundamental distinction between brown dwarfs and planets is the 
manner in which they are formed (e.g., Kaftos et al. 1986).  Brown 
dwarfs are formed in the same manner as a star, which means that 
if they are found as companions to stars then they formed by that 
process involved in the formation of a binary star system.  The underlying
mechanism in that case is not fully understood, but is thought to involve 
large-scale gravitational instabilities.  Planets, at least the nine that
we can study in detail, in contrast appear to be formed by an accretion 
process beginning from small dust grains building up to planetesimals 
to lunar-sized objects to terrestrial planet sized objects and then,
depending on the availability of gas, giant gas-rich planets such as Jupiter.
 
Thus, the distinction between brown dwarfs and planets is fundamental.
It is not a ``matter of semantics''.  An incorrect identification of 
the nature of these companions would lead in turn to erroneous notions 
about the basic processes involved in their formation and evolution.  
Importantly in this regard, as we do not at this time know either
the lower limit to the mass of a brown dwarf, or the upper limit to 
the mass of a planet, one cannot at this time use mass alone as the 
basis for identifying a given companion to a star as a planet, unless 
the masses are well below likely masses for brown dwarfs (i.e., Earth 
mass companions).
 
The sub-stellar mass companions discovered since 1989 are listed in Table 1.
There are two exceptions. The companion to HD 98230 was discovered, or at 
least the observations made, in 1931 (Bergman 1931).  Also, the Table 
lists only those companions discovered by radial velocity observations.  
The possible companion(s) to Lelande 21185 (Gatewood 1996) are not 
included as there is no reliable estimate at present of their eccentricity.

\clearpage
 
\begin{deluxetable}{crrr}
\footnotesize
\tablecaption{Properties of newly discovered substellar companions.  \label{tbl-1}}
\tablewidth{0pt}
\tablehead{
\colhead{STAR} & \colhead{MASS}   & \colhead{PERIOD (DAYS)}   & \colhead{ECCENTRICITY}
} 
\startdata
51 {Peg\tablenotemark{1}} &0.45 &4.23  &0.00 \nl
v {And\tablenotemark{2}}  &0.65 &4.61  &0.11 \nl
55 {Cnc\tablenotemark{2}} &0.84 &14.65 &0.05 \nl
Rho {CrB\tablenotemark{3}} &1.1  &39.64 &0.03 \nl
16 {Cyg\tablenotemark{4}} &1.6  &804   &0.65 \nl
47 {UMa\tablenotemark{2}} &2.3  &1090  &0.08 \nl
$\tau$ {Boo\tablenotemark{2}} &3.9 & 3.31 & 0.00 \nl
70 {Vir\tablenotemark{2}} &7.4 &116.7 &0.37 \nl
HD {114762\tablenotemark{5}} &9.0 &84.02 &0.33 \nl
HD {110833\tablenotemark{6}} &17 &270 &0.69 \nl
BD {-04:782\tablenotemark{6}} &21 &240.92 &0.28 \nl
HD {112758\tablenotemark{6}} &35 &103.22 &0.16 \nl
HD {98230\tablenotemark{7}} &37 &3.98 &0.00 \nl
HD {18445\tablenotemark{6}} &39 &554.67 &0.54 \nl
HD {29587\tablenotemark{6}} &40 &1471.70 &0.37 \nl
HD {140913\tablenotemark{6}} &46 &147.94 &0.61 \nl
BD {+26:730\tablenotemark{6}} &50 &1.79 &0.02 \nl
HD {89707\tablenotemark{6}} &54 &298.25 &0.95 \nl
HD {217580\tablenotemark{6}} &60 &454.66 &0.52 \nl
\enddata

 
\tablenotetext{1}{Mayor and Queloz (1995)}
\tablenotetext{2}{Butler \& Marcy (1996); Butler et al (1997); Marcy \& Butler (1996)}
\tablenotetext{3}{Noyes et al. (1997)}
\tablenotetext{4}{Cochran and Hatzes (1996)}
\tablenotetext{5}{Latham et al. (1989)}
\tablenotetext{6}{Mayor et al. (1996)}
\tablenotetext{7}{Bergman (1931)}
\end{deluxetable}
\clearpage

\section{An Orbital Signature of Formation Process}

The observational criterion suggested here for distinguishing between 
brown dwarf and planetary companions to a star concerns the relationship 
between two properties of the companion's orbit, namely its eccentricity 
as a function of its period.  Figure 1 shows orbital eccentricity 
versus the logarithm of orbital period, expressed in days, for two  
populations.  One population is pre-main-sequence (PMS) binaries
(Mathieu 1994). Another is objects thought to have formed via 
accretion in a disk, the giant planets in the solar system. Note 
that low eccentricity is a characteristic for other objects such as 
the terrestrial  planets, the regular satellites of the giant 
planets in the solar system, and the companions to the pulsar 
PSR 1257+12 (Wolszczan and Frail 1992), all of which are thought 
to have formed by accretion from a disk.

Data for PMS binaries were used as they presumably reflect any 
signature of the binary formation process with minimal alteration. 
Also, only systems with periods less than 10$^{4}$ days were used 
because of the apparent difference in the mass function for this
class of companions as compared with field stars and companions 
with periods longer than 10$^{4}$ days (Abt and Levy 1976, Mazeh et al. 
1992).  This difference has been taken as evidence by these 
authors and by Mathieu (1994) that the formation process for 
short-period binaries differs from that for field or long-period 
companions (see also Trimble and Cheung 1976, Trimble 1990).  
Main-sequence binaries with periods in the range considered here 
display a similar (e, log P) distribution as do the PMS binaries.  
Orbital evolution is thus not likely to be a major consideration 
for periods much beyond a few weeks.  Furthermore, it appears that 
the (e, log P) distribution is established early in the
history of these systems (Mathieu), consistent with a 
signature of their formation.

The distribution of (e, log P) for stellar companions is markedly 
different from that for planetary companions.  This difference is 
a reflection of a corresponding difference  in the manner by 
which objects in these two classes were formed.  A least-squares 
fit to the binary star distribution gives e = 0.2 log P - 0.03.  
This yields a Pearson correlation coefficient of r = 0.76.  
This statistical test suggest that the causal relationship between 
eccentricity and period is a modest but significant one. While a 
trend is clear, there is a range of eccentricities at a given period.

\section{A Test of the Criterion}

If the proposed (e, log P) criterion is valid one expects that 
the (e, log P) distribution for binary brown dwarfs would mimic, 
if not be indistinguishable from, that shown in Figure 1. Ten of 
the companions in Table 1 have 17 \lapprox M$_{c}$sin{\it i} \lapprox 
80 and as such are likely to be brown dwarfs.
 
The (e, log P) set for those ten companions have a Pearson coefficient 
r = 0.67.  This is similar to that for PMS binary companions.  The 
least-squares fit for the data gives e = 0.22 log P - 0.05, 
indistinguishable from that of PMS binaries given the uncertainties 
in the coefficients.
 
A better measure of the reality of these correlations is the Spearman 
coefficient as it is derived from a simple ranking of the data and 
does not depend on knowing the distribution function that might 
underlie the data (e.g., normal bivariate).  The Spearman coefficient 
for the brown dwarf companions is $\rho$ = 0.58.  This suggests that the
correlation between e and log P for this group of ten systems is 
significant at the 95 percent level.
 
It is clear from the above that the orbital signature of formation 
is the same for PMS binaries and the brown dwarf companions, 
demonstrating the validity of the criterion.

\section{Application of the Criterion}

We now apply the criterion to the group of substellar companions 
listed in Table 1 with M$_{c}$sin{\it i} \lapprox 10.  The least-squares fit 
gives e = 0.14 log P - 0.04.  This is similar to, but differs 
from the best fits to the PMS and brown dwarf populations.  
The Pearson coefficient for this group is r = 0.63, and the 
Spearman coefficient is $\rho$ = 0.73.
 
It should be noted that the results for this group are skewed heavily 
by two systems,  47 U Ma and Rho CrB.  If these two systems are 
ignored, and there is no a priori justification for so doing, the 
remaining systems yield a best fit of e = 0.23 log P - 0.10. 
This gives values of e at a given value of log P that are in good 
agreement with the values calculated from the best-fit lines for PMS 
and brown dwarfs.  The Pearson coefficient for this fit is r = 0.95, 
and the Spearman coefficient is $\rho$ =  0.96.  This trend is both 
statistically significant and consistent with that of the 
PMS binaries and brown dwarfs.

\section{Discussion}

Shown in Figure 2 are all three populations, PMS, brown dwarfs, 
and companions  with M$_{c}$sin{\it i} \lapprox 10.  Also shown for reference 
is the best fit line for the combined population. The Spearman  
coefficient for the three  populations combined is 0.75.  With
a sample consisting of 44 members, this value of the Spearman 
coefficient indicates that the observed correlation is significant 
at greater than the 99.999 percent confidence level.

It would appear that with two exceptions, a third if we include 
the astrometric system LeLande 21185, all of the newly discovered, 
sub-stellar mass companions, have an (e, log P) distribution 
that could be drawn from the same population and that this distribution 
is indistinguishable from that of PMS binary companions.  Given that this  
distribution is a signature of their formation process, it would appear 
that these companions formed by the same process as did the binary 
stellar companions.  That is, those companions display the orbital 
signature of brown dwarfs and not that of planets.
 
Some authors (e.g., Butler et al.) have suggested that 
companions with low eccentricity, such as the putative 
companion to 51 Pegasi, are planets by virtue of their 
low eccentricity.  However, as is clear from Figure 1 and 
knowledge of tidal effects (e.g., Duquennoy and Mayor 1992), 
any companion that is that close to a star will be in a low  
eccentricity orbit. The substellar companions with short 
periods are fully consistent with them being members of 
the stellar population.  For any companion that has a sufficiently
short period that its orbit will be tidally circularized, it's 
eccentricity alone cannot be used as a criterion for identifying 
it as a  planet.

There is an additional orbital signature that could be 
useful as a criterion on whether a companion is a planet 
(Black 1980).  This arises again from the formation  mode, 
namely, more than one companion is formed and the resultant 
orbital architecture is characteristic of that process, 
and that architecture differs from that of a multiple star
system.  The regular satellite systems of Jupiter, Saturn, 
and Uranus, as well as the companions to the pulsar PSR 
1257+12, and the planets in the Solar System are all multiple 
systems with geometric spacing of the orbits.  The five 
systems that we have reason to believe have companions 
formed by accretion all have the same basic architecture.  
If the companions that have been discovered are planets, 
one can expect that other companions are present.  So far, 
no such companions have been found.  Their continued absence 
would weaken the interpretation of the companions in Table 1 as       
planets, whereas detection of such companions strengthens that 
interpretation.

Nature knows how to make even stellar mass companions with 
short orbital periods, so if these objects are brown dwarfs, 
their presence close to the star is not a mystery, but is to 
be expected.  Also, the high eccentricities that are seen 
for the putative planets with orbital periods longer than a 
few weeks is perfectly natural if they are brown dwarfs.  
For example, given the orbital period of the companion to 
16 Cyg B, the best fit line {\it based just on the PMS binary 
star distribution} would predict an eccentricity of  0.55
for the companion.  The observed eccentricity is 0.65, 
good agreement given the uncertainty in both the fit and 
the data. Similarly, the predicted eccentricity for the 
companion to 70 Vir is 0.38,  virtually identical with the 
observed value of 0.37. The fit is even better if one uses 
the regression defined by all the data points shown in Figure 2.
There is no need to postulate a form of  dynamical instability 
arising from the presence of a distant binary to account for 
the observed eccentricity (e.g., Holman et al. 1997, Innanen
et al 1997).  Indeed it is problematic whether the proposed 
mechanism would work in the presence of a system where there 
are other sources of gravity such as other companions, as would 
be the case if the companion is a planet, or a disk, in the 
vicinity of the companion to 16 Cyg.  This caveat is pointed 
out by both Holman et al. and Innanen et al.  There is no binary  
in the case of 70 Vir to cause the  eccentricity of this companion
so if it is a ``planet'', one must invoke yet another mechanism.  
The binary formation mechanism would appear to provide a single, 
unifying context for the observed orbital eccentricities and their 
systematic variation with orbital period.

The interpretation of the nature of the newly discovered 
substellar companions that emerges from the (e, log P) 
criterion also removes the need for mechanisms (Lin et
al. 1996, Trilling et al. 1996, Weidenschilling and Marzari 
1996) that postulate significant (i.e., nearly unity 
fractional change in semi-major axis) orbital evolution to
account for short-period companions.  It should be noted 
that contrary to statements elsewhere (e.g., Marcy and 
Butler) the location of the apparent companion to 47
U Ma has not been shown to be inconsistent with where a 
giant planet might form based upon our current paradigm for 
gas giant planet formation.  The companion to Rho CrB,
if it is a planet, would likely have experienced orbital migration, 
but an alternative interpretation is that it is a brown dwarf, 
formed near its current location, with a slightly lower eccentricity 
than is typical for such companions with that orbital period.

This does not mean that such mechanisms cannot, or do not, 
occur in planetary systems, but there is no independent 
evidence that such large-scale migration occurs in those 
disks that end up making planetary systems.  On the 
contrary, the five bona fide systems where it is believed 
that accretion-formed companions exist all show no signs
(e.g., eccentric and non-geometric spaced orbits) of such 
motion.  It will be important to understand the {\it testable}, 
that is observable, consequences of models that suggest that
systems such as 51 Peg are the result of large-scale motion 
through a disk.  Clearly one such observable test for 
those models that rely upon gravitational scattering of one giant
planet by another would be the presence of a second companion 
in a very high eccentricity, longer period orbit.  Failure 
to find such companions would rule out such models.  Models 
that rely solely on interactions between a single large planet 
and its parent disk are more difficult to verify by examining 
only their end state.  Their verification will likely await 
instruments capable of very high spatial resolution far-infrared 
to sub-millimeter studies of young systems where the possible 
detection of density waves in such disks could signal that 
the process in question is taking place.

Another observational criterion for distinguishing between 
brown dwarfs and planets was suggested originally by Lunine 
(1986) and refined recently (Saumon et al.  1996).  The basic 
notion is that as planets form via accretion with solid material 
playing a key role, one expects that the abundance of metals 
relative to hydrogen will be greater than in the central star.  
That is the case for Jupiter relative to the Sun.  In contrast, 
as a brown dwarf is formed by the same process as a star, one 
expects that the composition of a brown dwarf companion will be 
similar to, if not identical with, that of its stellar companion.  
A  spectroscopist confronted with two Jupiter-mass objects, 
one a planet and the other a brown dwarf, should be able in 
principle to distinguish which is which on the basis the metal 
to hydrogen ratio normalized to the same ratio in the central star.
Spectroscopic studies of companions to other stars are beyond 
the capability of existing instruments except for situations 
where the companion is both relatively luminous and well-separated 
from its stellar companion.  Thus, while this observational test
has potential, its full application must also await future 
instrumentation.

The perspective offered here raises the important point that 
these new detections are providing the first observational guide 
to the lower limit to the mass of a brown dwarf.  Assuming that 
there is nothing pathological in the viewing geometry for these   
systems (i.e., sin{\it i} $\sim$ 0), then it would appear that the lower 
limit is comparable to the mass of Jupiter.  It will be 
interesting to see whether even less-massive companions are     
discovered with short orbital periods.  Systems with the 
accuracy of those now in use should be able to detect companions 
with masses as low as a few tens of Earth masses in orbits with 
periods of a few days.

{\bf Acknowledgement: }	I wish to thank the referee for useful comments on the manuscript.  This research 
was conducted at the Lunar and Planetary Institute which is operated by the Universities 
Space Research Association under contract NASW-4574 with the National Aeronautics 
and Space Administration.

\clearpage

\begin{figure}
\plotone{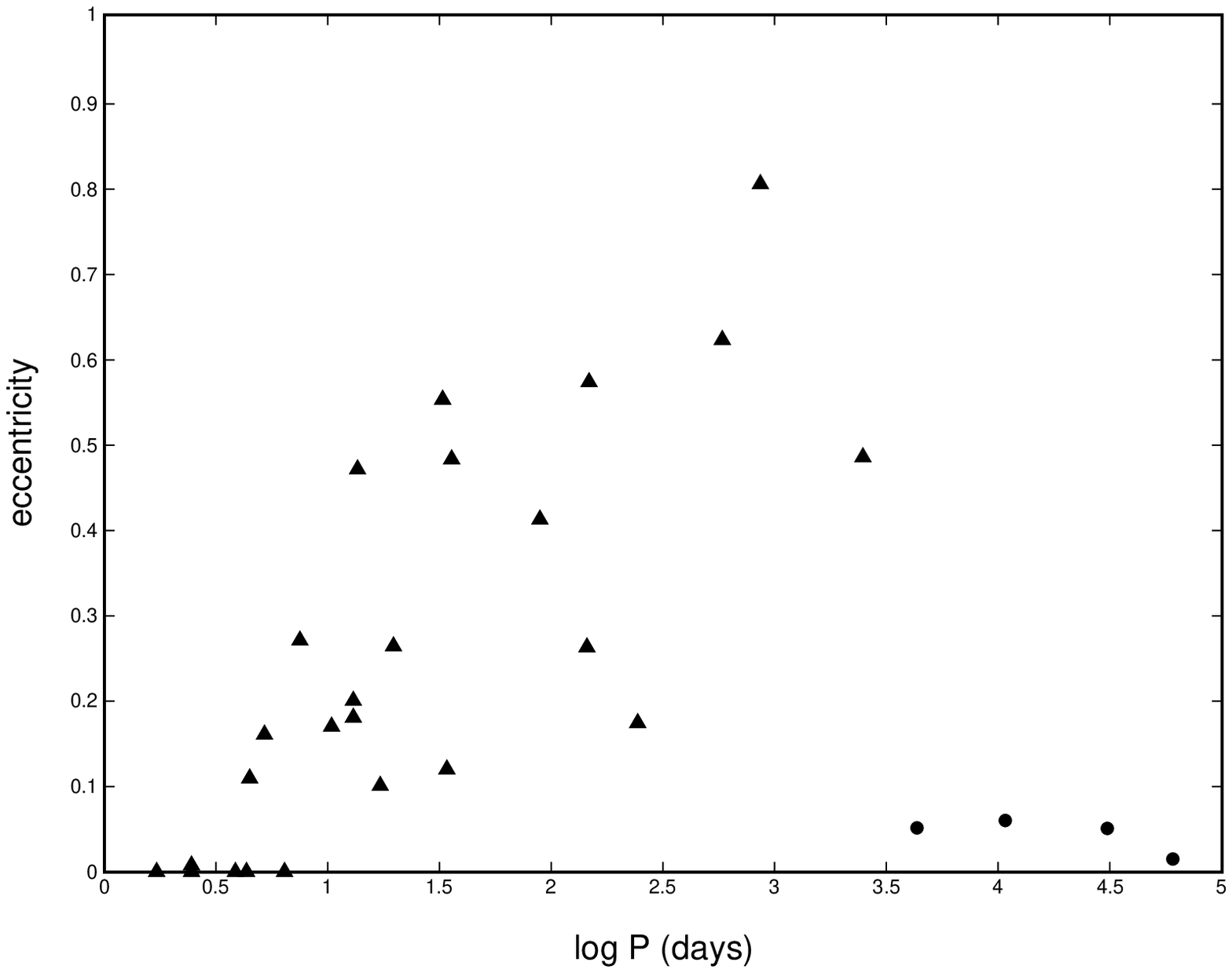}
\caption{Orbital eccentricity as a function of orbital
period in days for two populations, pre-main sequence binary stars
(filled triangles; Mathieu, 1994) and giant planets in the solar system
(filled circles). \label{fig1}}
\end{figure}
\clearpage

\clearpage
\begin{figure}
\plotone{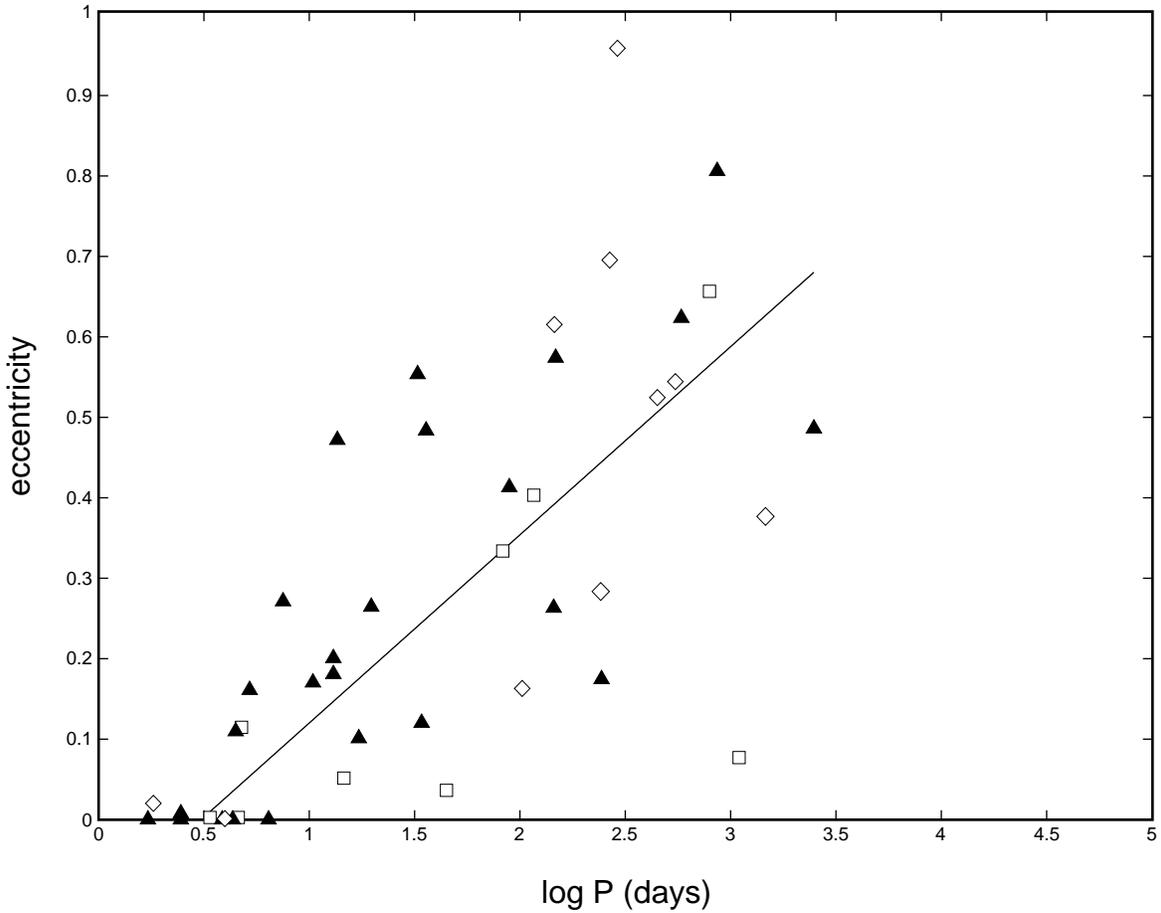}
\caption{Same as Figure 1, but for three populations.
Pre-main sequence binaries (filled triangles), companions with minimum
masses greater than ten but less than eighty Jupiter masses (open diamonds;
Mayor et al. 1996), and companions with minimum masses less than ten Jupiter
masses (open squares; Latham et al 1989, Mayor and Queloz 1995, Marcy and
Butler 1996).  Also shown for reference is the best-fit line for the
combined populations. \label{fig2}}
\end{figure}
\clearpage

\end{document}